# Differentiation of skin incision and laparoscopic trocar insertion via quantifying transient bradycardia measured by electrocardiogram

Cheng-Hsi Chang, Yue-Lin Fang, Yu-Jung Wang, Hau-tieng Wu and Yu-Ting Lin

From the Department of Anesthesiology (CC,YW) and Department of Surgery (YF), Shin Kong Wu Ho-Su Memorial Hospital, Taipei, Taiwan; Department of Mathematics and Department of Statistical Science, Duke University, Durham, NC, U.S.A. (HW); Department of Anesthesiology, Taipei Veteran General Hospital, Taipei, Taiwan. (YL); College of Medicine, Fu Jen Catholic University, New Taipei, Taiwan (CC).

* Corresponding Author: Yu-Ting Lin. Address: Department of Anesthesiology, Taipei-Veterans General Hospital, 201, Section 2, Shih-Pai Road, Taipei, Taiwan, Telephone number: 886-2-28757549, Fax number: 886-2-28751597, E-mail: linyuting@hotmail.com.tw

# Abstract

**Background.** Most surgical procedures involve structures deeper than the skin. However, the difference in surgical noxious stimulation between skin incision and laparoscopic trocar insertion is unknown. By analyzing instantaneous heart rate (IHR) calculated from the electrocardiogram, in particular the transient bradycardia in response to surgical stimuli, this study investigates surgical noxious stimuli arising from skin incision and laparoscopic trocar insertion.

**Methods.** Thirty-five patients undergoing laparoscopic cholecystectomy were enrolled in this prospective observational study. Sequential surgical steps including umbilical skin incision (11 mm), umbilical trocar insertion (11 mm), xiphoid skin incision (5 mm), xiphoid trocar insertion (5 mm), subcostal skin incision (3 mm), and subcostal trocar insertion (3 mm) were investigated. IHR was derived from electrocardiography and calculated by the modern time-varying power spectrum. Similar to the classical heart rate variability analysis, the time-varying low frequency power (tvLF), time-varying high frequency power (tvHF), and tvLF-to-tvHF ratio (tvLHR) were calculated. Prediction probability ($P_K$) analysis and global pointwise *F*-test were used to compare the performance between indices and the heart rate readings from the patient monitor.

**Results.** Analysis of IHR showed that surgical stimulus elicits a transient bradycardia, followed by the increase of heart rate. Transient bradycardia is more significant in trocar insertion than skin incision. The IHR change quantifies differential responses to different surgical intensity. Serial $P_K$ analysis demonstrates de-sensitization in skin incision, but not in laparoscopic trocar insertion.

**Conclusions.** Quantitative indices present the transient bradycardia introduced by noxious stimulation. The results indicate different effects between skin incision and trocar insertion.

# Introduction

Surgical noxious stimulation elicits dynamic response in human body. Recent development of monitoring instrument has provided us various methods to assess noxious stimulation and analgesia. For example, the surgical plethysmographic index (SPI) and analgesia nociception index (ANI), which use autonomic responses to measure noxious stimulation.[1–4] Response Entropy[TM] (RE) and Bispectral index (BIS) provide information from processed electroencephalography and electromyography.[5–8] Besides, we also use the readings of heart rate (HR) and blood pressure (BP) from standard patient monitor.[9] Many researches provides a wealth of knowledge regarding the noxious stimulation arising from the skin[1,3,5,9,10,11] while surgical procedures usually elicit deeper structure stimulation after the initial skin incision. The surgical noxious stimulation of deeper structures is clinically more important than that of the skin.

It is known that superficial cutaneous pain is different from that of deeper structures, such as the muscle, joint and visceral organ.[12] Cutaneous pain is perceived as sharp intense sensation and accompanied with autonomic responses of tachycardia, hypertension and a fight-or-flight behaviour. On the other hand, pain arising from deeper structures of human body is more characterized as an aching sensation, causing bradycardia, hypotension and hypoactive behaviour. At the extreme of another end, visceral pain arising from internal organ is often described as vague in location.[12–14] However, to obtain this pain quality description from an anaesthetized patient may need a new approach.

Our preliminary study[15] of *instantaneous heart rate* (IHR) showed that noxious stimulation elicits a transient bradycardia lasting for few seconds, before the well-known tachycardia. This transient change of HR can be observed from the R-to-R peak intervals (RRI) from electrocardiography (ECG). Unfortunately, the HR reading from the patient monitor might not faithfully present the transient bradycardia, mainly because the HR reading is based on "counting heart beats" within a time window, which neutralizes the opposing effects of increase and decrease on heart rate. Through IHR, we may peek into different autonomic profiles during surgical noxious stimuli.

We hypothesized that different surgical steps in different sizes and tissue types, on cutaneous or deep structures, cause different dynamic changes in IHR. Using prospective observation of standard laparoscopic cholecystectomy surgery, we aimed first to investigate the surgical effects between umbilical, xiphoid and subcostal areas, which were different in size. Second, we aimed to compare the difference between skin incision and laparoscopic trocar insertion, which were different in histological tissue type.

# Methods

We conducted a single center, prospective, observational cohort study at Shin Kong Wu Ho-Su Memorial Hospital, Taipei, Taiwan to investigate the IHR during surgical noxious stimuli on different location and at different intensity. The study was approved by the institutional ethics committee (Institutional review board of Shin Kong Wu Ho-Su Memorial Hospital, Taipei, Taiwan; Address: No. 95, Wen Chang Road, Shih Lin District, Taipei, Taiwan; Protocol No.: 20160706R;Chairman: Gong-Jhe Wu, M.D., Ph.D.) on 10 November 2016, and written informed consent was obtained from each patient. From Dec. 2016 to Oct. 2017, we enrolled 35 patients, ASA I to III, scheduled for laparoscopic cholecystectomy because of medical conditions such as cholecystitis, gall stone or gall bladder tumor. All laparoscopic cholecystectomy (LC) surgeries were performed by the same surgeon (the author Y.F.) to provide standard and consistent surgical procedure while the physiological data including electrocardiography (ECG) were recorded continuously as well as the precise timestamps of the events. Exclusion criteria include major cardiac problem, uncontrolled hypertension, arrhythmia shown in pre-operative ECG, known neurological disease, history of drug abuse and anticipated difficult airway.

## Study protocol

After at least eight-hour fasting, the patients were sent in the operating theatre. Standard patient monitor (Philips IntelliVue$^{TM}$ MP60), including ECG, pulse oximeter (SpO$_2$) and non-invasive blood pressure, were applied to all patients. A BIS electrode (BIS-XP sensor) was applied to patients before anaesthetic induction. As our usual practice, all the anaesthetic managements, including the administration of medications, and airway management were under the discretion of the anaesthetist. After intravenous infusion of lactated Ringer's solution was started, we started anaesthetic induction with fentanyl 2.5 – 3 μg/kg;

after preoxygenation, hypnosis was induced with propofol 2 – 2.5 mg/kg. Rocuronium was used to facilitate tracheal intubation. Subsequently, mechanical ventilation was started from volume control mode (tidal volume 7 ml/kg) with oxygen-gas-sevoflurane mixture as low flow (<1L) anaesthesia. The adequacy of anaesthetic depth (during the whole period of surgery) was determined by BIS index (<65) and the anaesthetist's judgments as the usual. Bolus dose of fentanyl was given if inadequate analgesia was determined. Atropine or other anticolinergic medication was only given for marked bradycardia (HR<40 bpm).

The surgical steps were as follows. After disinfection and draping, the surgery was started with umbilical skin incision (11 mm) made by the surgeon using a scalpel (handle: 3#, blade: 15#). Then a Veress needle (Sigma Medical Supplies Corp., Taipei, Taiwan) was inserted via the umbilical skin wound for $CO_2$ insufflation (pressure 12 mmHg). After achieving pneumoperitoneum and reverse Trendelenburg's position, a laparoscopic trocar with tip (Kii Optical Access system CTR33, Applied Medical, CA, USA) was inserted via the umbilical wound. Subsequently, skin incision in 5 mm was made at xiphoid area using the same instrument, and the second laparoscopic trocar insertion with tip (Kii Optical Access system CTR03, Applied Medical, CA, USA) was inserted via the xiphoid wound. Finally, a 3 mm subcostal skin incision was made as well as a 3 mm in diameter trocar was inserted to establish all laparoscopic ports. We focused on the following sequential surgical steps for data analysis: umbilical skin incision (11 mm), umbilical trocar insertion (11 mm), xiphoid skin incision (5 mm), xiphoid trocar insertion (5 mm), subcostal skin incision (3 mm), and subcostal trocar insertion (3 mm). All surgeries followed this standard sequence. Intervals of consecutive steps were one minute apart at least. The size of surgical wound is standardized by the standard size of surgical instruments and carried out by a single surgeon.

## Data collection

The physiological waveform data including ECG were recorded continuously from the patient monitor (Philips IntelliVue[TM] MP60) using a data acquisition software (ixTrend Express, ixellence, Wildau, Germany) running on a laptop. The sampling rates of EASI mode ECG lead is 500Hz. At the moment of each surgical step, the exact timestamp was registered by using a purpose-made software so that a one-click movement can register the surgical events and its accurate time. Data of the gas analyzer (GE Datex-Ohmeda S5 Compact) was recorded using the data acquisition software, Datex-Ohmeda S/5 collect ( GE Health Care Finland Oy, Helsinki, Finland). The offline ECG data were analyzed by the following steps to extract IHR information. Standard R peak detection algorithm was automatically performed on EASI ECG lead and confirmed by visual inspection. The ectopic beats were removed and the R-to-R interval (RRI) was interpolated using cubic spline interpolation. IHR is the inverse of RRI in beat-per-minute (bpm) unit. We excluded IHR of insufficient quality. The criteria of optimal IHR are that the maximum should be less than 120 bpm and the minimum should be larger than 40 bpm.

# IHR quantification

To quantify the transient bradycardia, we used a time-varying power spectrum (tvPS) technique developed in our previous study.[15] It is in essence a high resolution time-frequency (TF) analysis referred to as *concentration of frequency and time* (ConceFT).[16] In brief, ConceFT integrate the benefit of the multitaper technique and the nonlinearity of a chosen TF analysis to achieve high resolution, with which we could quantify the transient dynamics inside the IHR signal, particularly in the spectral sense.[17][18] The theoretic ground is still the same as the classical HRV analysis: high frequency power represents vagal activity, low frequency power represents sympathetic activity, and the power spectrum provides frequency information under stationary condition.[19] ConceFT simply removes the restriction of stationary condition so that we can explore the transient IHR change from the patient undergoing surgery. The main quantities we consider are the time-varying low frequency power (tvLF), time-varying high frequency power (tvHF), and tvLF-to-tvHF ratio (tvLHR), which are evaluated from the tvPS.

Inter-individual variability is corrected by normalizing the IHR by the median heart rate during the 120 s period before intubation. The ordinary heart rate (OHR) is the direct read-out HR data from the patient monitor. The same normalization procedure is also applied to OHR to correct the inter-individual variability.

We optimized the detection performance of ConceFT on our previous database[15] comprising 59 cases as the training dataset. Particularly, the window length (60s) and the number of ConceFT (K=30) were trained to obtain the optimal performance of transient bradycardia detection. Second, we applied the algorithm with trained parameters to dataset collected in this prospective observation study comprising 35 cases to quantify the transient IHR change. This validation step confirms the robustness of the proposed algorithm.

## Statistics

Considering the unknown quantitative change of IHR caused by trocar insertion, and the inter-individual variation, we estimated the normalized IHR and its standard deviation to be 1±30% and an observable IHR change with 15% mean change will lead to a sample size larger than 31 for adequate study power, given a 5% type I error and a 20% type II error. A case number of 35 adds additional margin in statistics.

Prediction probability ($P_K$) analysis developed by Smith et al.[20] is applied as the statistical performance measurement of indices in anaesthesia. Serial $P_K$ ($sP_K$) analysis has been used to reveal the performance and the temporal relationship to distinct anaesthetic events.[6,15,17] Practically, we calculated the $P_K$ value and its standard error by using Jackknife method. Baseline was set as 20 s before each surgical step. By plotting the $sP_K$ value and its standard error bar, we can do hypothesis test simply with the naked eye. The significant difference (P < 0.05) between indices can be established if 1.5 times of their standard error bars do not overlap.[17,20] In addition to the above three time-varying indices, we added IHR and OHR as indices in the $sP_K$ analysis.

The IHRs of all subjects around different surgical events are summarized with the functional mean and one standard deviation. To enhance the visualization, the IHRs around different surgical events are further normalized by the median of IHR over a 30 s interval, from −40 s to −10 s, prior of the event. Similarly, the tvHF, tvLF, and tvLHR are summarized with the functional mean and one standard deviation. To compare if tvLF (tvHF, tvLHR, and OHR respectively) reflects different noxious stimulation, we apply the global pointwise $F$-test (GPF)[21] on tvLF (tvHF, tvLHR, and OHR respectively) over the 20-s period after the event on different surgical events. The p-value less than 0.05 is viewed as statistical significant. Besides, all statistical results are expressed as mean (standard deviation). All data analysis and statistics is performed with MATLAB software (MATLAB R2016b, MathWorks, Natick, MA, USA). Finally, all above statistical analyses were performed within short time duration of data around each surgical step (−40 s to +40 s) to minimize the influence from outside that period.

# Results

Thirty-five cases (21 women and 14 men) were enrolled within the study period to provide the data for analysis. Their mean age was 48 (11.4) yr (range 26–72 yr) and BMI was 27(5.1) kg m$^{-2}$ (range 20.7–45.2). Data with sufficient quality collected during umbilical skin incision, umbilical trocar insertion, xiphoid skin incision, xiphoid trocar insertion, subcostal skin incision, and subcostal trocar insertion are 35, 34, 29, 29, 27, and 26 respectively. Anticholinergic drug was not given during the period of these data recording.

The observation of data (Fig.1) and the group average of IHR (Fig.2) showed that surgical stimulus elicits a transient bradycardia, followed by the increase of HR. Transient bradycardia appears more significant in trocar insertion than skin incision. The IHR change also appears more significant in trocar insertion than skin incision. For skin incision, the subsequent xiphoid skin incision and subcostal skin incision barely elicited IHR change than the first umbilical skin incision. In contrast, all three trocar insertions elicit obvious transient bradycardia, which followed by the tachycardia in response to all surgical stimulation.

Comparison of indices including tvHF, tvLF and tvLHR (Fig. 3) showed that on average all three trocar insertions elicit stronger responses than umbilical skin incision. For tvHF and tvLHR, the GPF functional hypothesis testing (Table 1) shows a statistical difference among umbilical skin incision and three trocar insertions (p<0.001 and p=0.028 respectively). For the two groups test, the functional hypothesis testing shows a statistical difference between xiphoid trocar insertion and umbilical skin incision (p=0.021 and p=0.014 respectively). Besides these combinations, no statistical difference was found.

In the one-minute time window of, $P_K$ values of Indices show significantly better predictive ability than the ordinary heart rate value in first skin incision (tvLF>0.82, tvHF>0.58, tvLHR>0.80, OHR<0.52) and all three laparoscopic trocar insertion (Umbilical: tvLF>0.86, tvHF>0.76, tvLHR>0.84, OHR<0.54; Xiphoid:

tvLF>0.86, tvHF>0.80, tvLHR>0.88, OHR<0.58; Subcostal: tvLF>0.84, tvHF>0.70, tvLHR>0.86, OHR<0.56). Also only the first skin incision (umbilical) elicits significantly high $sP_K$ values (>0.8) of tvLF and tvLHR. The $sP_K$ values in subsequent xiphoid and subcostal skin incision cannot reach more than 0.7 for all indices. For all three trocar insertion stimuli, the $sP_K$ values of tvLF, tvHF and tvLHR showed good response (>0.7). The tvHF has better $sP_K$ performance in trocar insertion than that in skin incision. Also, tvHF has earliest response among other indices in trocar insertion. In addition, treating IHR per se and OHR as indices yields no good $sP_K$ performance (<0.6) within 40 s period after noxious stimulus.

# Discussion

There are two major results from this prospective observational study. First, the transient IHR change contains abundant quantitative measurement on the instant of surgical stimuli, in contrast to the ordinary heart rate reading of the patient monitor. Second, the laparoscopic trocar insertion elicits stronger HR change and stronger transient bradycardia compared to skin incision at the same surgical site and size. It indicates that IHR contains microscopic time scale information reflecting the surgical noxious stimulation of skin and deeper structure, and their differences.

    The normalized and group-averaged IHR (Fig. 2) showed a typical behaviour that transient bradycardia could occur immediately after noxious stimulation followed by the marked tachycardia, in particular the first skin incision and all trocar insertions. In $sP_K$ analysis (Fig.4), tvHF also reflects the first response of transient bradycardia to stimulus. We might be aware of the sympathetic activation and overlook the parasympathetic activation by looking at the standard patient monitor since its HR reading is the averaged product of IHR. Therefore, our findings are consistent with other studies and common clinical experience that skin noxious stimulation elicits increase of HR.[3 5 7] The group-averaged IHR also showed large variance. It is a direct result of inter-individual variability and the complex surgical effect on human body, as well as the wide age and BMI range of this patient group. As the surgical steps goes from umbilical area, xiphoid area to subcostal area, the wound size goes from 11mm, 5mm to 3mm. Compared with the first umbilical skin incision, the subsequent surgical stimuli elicited less consistent increase of HR, which indicates HR became less and less useful for those subsequent surgical stimuli. As for the transient IHR changes, the subsequent skin incisions produced trivial response of tvHF and tvLF, indicating physiological de-sensitization. In contrast, the decrease of trocar insertion intensity did not impair the performance of tvHF and tvLF indices, which can be judged from the side-by-side comparison and $sP_K$ analysis (Fig. 3, 4). This result

showed that de-sensitization is also different between skin incision and trocar insertion.

Laparoscopic cholecystectomy is a representation of the modern trend of minimally invasive surgery. Strictly speaking, both skin incision and trocar insertion produced surgical wound sizes no larger than 11mm in size, which are not comparable to the surgical stress intensity of traditional major surgery. Therefore, it is probably safe to say that the laparoscopic surgery provides us unified and discrete surgical stimulation to observe the IHR change. In this study, the umbilical wound is at the anatomical midline while the xiphoid wound is located at slight rightward off-midline near the xiphoid process. Therefore, the xiphoid trocar insertion includes the penetration of a muscle layer while the umbilical trocar insertion does not (only the linea alba). This anatomical structure difference might explain the less than expected IHR response of umbilical trocar insertion despite the largest wound length (11mm) and the statistical non-significance. The comparable sizes and locations of three surgical locations support the finding that noxious stimulation arising from the skin is different from that of the deep structure including muscle and peritoneum.

It is important to discuss the terminology of "transient bradycardia". In literature, bradycardia is commonly referred to as a marked slowing of heart rate lower than 60 bpm, usually accompanied with violent clinical symptoms, such as syncope or shock.[22][23] Some surgical procedures are well known for causing such bradycardia. For example, the oculocardiac reflex is known as a surgical stimulation around the eyeball eliciting marked bradycardia or even asystole, which is presumed to be vagally mediated from the central nerve system.[24][25] In addition, vagal activation induced by laparoscopic surgical manipulation could be the cause of bradycardia and cardiovascular collapse.[26] In contrast, we generalize the notion in this study and refer the subtle HR slowing after noxious stimulus to as "transient bradycardia".

Transient bradycardia in response to noxious stimulus has been presented in literature. Seitsonen et al. reported a study regarding measurement responses of various modalities to noxious stimulation.[5] The transient bradycardia was shown as group-averaged RRI that lasts for less than 10 s judging from their second figure.[5] Ekman et al. have also reported an initial slowing of HR immediately after laryngoscope, followed by increased HR in some cases (10/21).[7] It also appears at endotracheal intubation in pediatric patient.[8] Animal study[27,28] showed that noxious stimulation elicits a simultaneous activation of both sympathetic and parasympathetic nerve system. However, the increase of HR usually overpowers transient bradycardia so that the transient bradycardia is difficult to identify in the standard patient monitor since its HR reading is an average value of heart beats in 10-s window.[15,29] Since we investigated the trocar insertion in this study and found the transient bradycardia occurs also and being more obvious than that from skin incision, to our best knowledge, this is the first quantitative result of transient bradycardia in response to surgical stimuli in tissues deeper than skin, which should carry more clinical meaning.

The laparoscopic trocar penetrates muscle layer and parietal peritoneum while the scalpel cuts the skin. Their different responses to noxious stimuli indicate distinct physiological mechanisms. The characteristics of cardiovascular, behaviour response and neuro-physiological mechanism between cutaneous pain and deep somatic pain are different in function and anatomy.[12–14] Cutaneous pain evokes hypertension, tachycardia, protective reflex and increased alertness, whereas pain arising from deep somatic structure (muscle, joint) and visceral organs produces vague localization, hypotension, slowing pulse rate and hypo-reactive behaviour. Opioids exert analgesic effect at the midbrain periaqueductal gray region (PAG). Animal studies have reported that cutaneous pain activates lateral periaqueductal gray matter (lPAG), which mediates hemodynamic, autonomic and behaviour responses as fight-or-flight, while deep tissue pain activates ventrolateral periaqueductal gray matter (vlPAG) mediating the contrary responses.[30–34] The similar topographic distribution of neural mechanism regarding pain quality and hemodynamic connection also have been reported from human studies.[35,36] In physiology, the large amount of evidence has solidified the two types of pain: the fast pain and the slow pain,[14] which provides us the theoretical ground to take advantage of the difference between their autonomic responses.

The deep tissue noxious stimulation is more relevant to clinical anaesthesia since most surgical procedure is not only carried out at skin deep. Hence, we employ the standard laparoscopic surgery to profile the trocar insertion. The measurement of deep tissue pain may provide information additional to the readings of HR and BP during anaesthesia because hypotension or bradycardia might not reassure adequate analgesia. SPI and ANI are monitoring instruments dedicated to provide indices for the state of analgesia.[1,3] While SPI measures the change of heart rate and finger vascular flow, ANI compute RRI to gain quantitative information. Both indices provide information regarding the state of analgesia, in particular, a latent status of the balance between noxious stimulation and the analgesia to guide the administration of analgesics.[3][4] Broadly speaking, both ANI and the transient HR in this study is a kind of HRV analysis. The methodology of ANI is a measurement of the coherence of the respiratory influence over a period of time to assess the states of analgesia/nociception balance.[4] In contrast, the transient HR change possesses the dynamic nature, presented as the short duration performance in the $sP_K$ analysis. Hence, it is not currently suitable for the role of "latent state" indices. How to incorporate the transient HR information into the assessment of surgical noxious stimulation will be an important research in the future.

There are other limitations of this study. First, the main limitation is that the epidemiology of our case population is not homogeneous. The large range of age and BMI increases the inter-individual variability, even with a normalization of the IHR before analysis. The surgery also exerted influences on human body, such as pneumoperitoneum and position change. This limitation, however, enhances the strength — the transient IHR change still stands out from the data analysis. Second, the one-minute interval between each consecutive surgical step prevents us from observing the surgical noxious effect for longer than one minute, which interfere the physiological effect of subsequent noxious stimulation also.

Nevertheless the clinical value of transient IHR is well within the one minute period according to the results in this study (Fig. 1, 3, 4) and our previous study.[15] Third, the result cannot be applied in our clinical practice easily. The transient IHR change requires computation from ECG waveform, but the current standard patient monitor does not provide this functionality. Forth, laparoscopic surgery is a kind of minimal invasive surgery. Hence, our result maybe not representative to other surgeries requiring larger wound and stronger surgical stress. The wider types and locations of surgical manipulation, features of the corresponding noxious stimulus, and their merits on anaesthetic management need to be investigated. Fifth, the performance of indices, such as tvLF and tvHF are good in presenting the surgical noxious stimulation, but only for a short period around 20 s. Therefore, these indices provide dynamic information, rather than the state of analgesia/nociception balance to guiding the management of analgesia, which is the main purpose of the current analgesia monitors. How to incorporate the dynamic information into the assessment of analgesia state should be investigated in the future. Finally, cardiac rhythm other than sinus rhythm precludes the observation of transient bradycardia from IHR.

In summary, we quantified the transient IHR change, particularly the transient bradycardia introduced by surgical noxious stimulation. The results indicate differential autonomic responses between skin incision and laparoscopic trocar insertion, and the responses of surgical intensity.

# Acknowledgements

Assistance with the study: none.

Sponsorship: The work of C.-H. Chang was supported by the institutional research grant (SKH-8302-105-DR-20) from Shin Kong Wu Ho-Su Memorial Hospital, Taipei, Taiwan. The work of Y.-T. Lin was supported by the National Science and Technology Development Fund (MOST 106-2115-M-075-001) of Ministry of Science and Technology, Taipei, Taiwan.

Conflicts of interest: none.

Presentation: The preliminary work has been presented in poster display of the 2017 Canadian Anesthesiologists' Society (CAS) Annual Meeting held at June 23-26, 2017 in Niagara Falls, ON, Canada, and in poster display of the 2018 Society for Technology in Anesthesia (STA) Annual Meeting held at January 10-13, 2018 in Miami, FL, U.S.A.

# Table

**Table 1.** Significance *p* values of global pointwise F-test method

|                        | LF    | HF      | LHR    |
|------------------------|-------|---------|--------|
| UI vs. UT              | 0.613 | 0.117   | 0.138  |
| UI vs. XT              | 0.290 | 0.021*  | 0.014* |
| UI vs. ST              | 0.629 | 0.286   | 0.110  |
| UI vs. UT vs. XT vs. ST| 0.595 | <0.001* | 0.028* |

Significance *p* values of functional hypothesis testing of indices by the 20-s period after each surgical step using global pointwise F-test method. LF, time-varying low frequency power; HF, time-varying high frequency power; LHR, LF-to-HF ratio; UI, umbilical skin incision; UT, umbilical trocar insertion; XT, xiphoid trocar insertion; ST, subcostal trocar insertion; *, p<0.05.

# Figure Legends

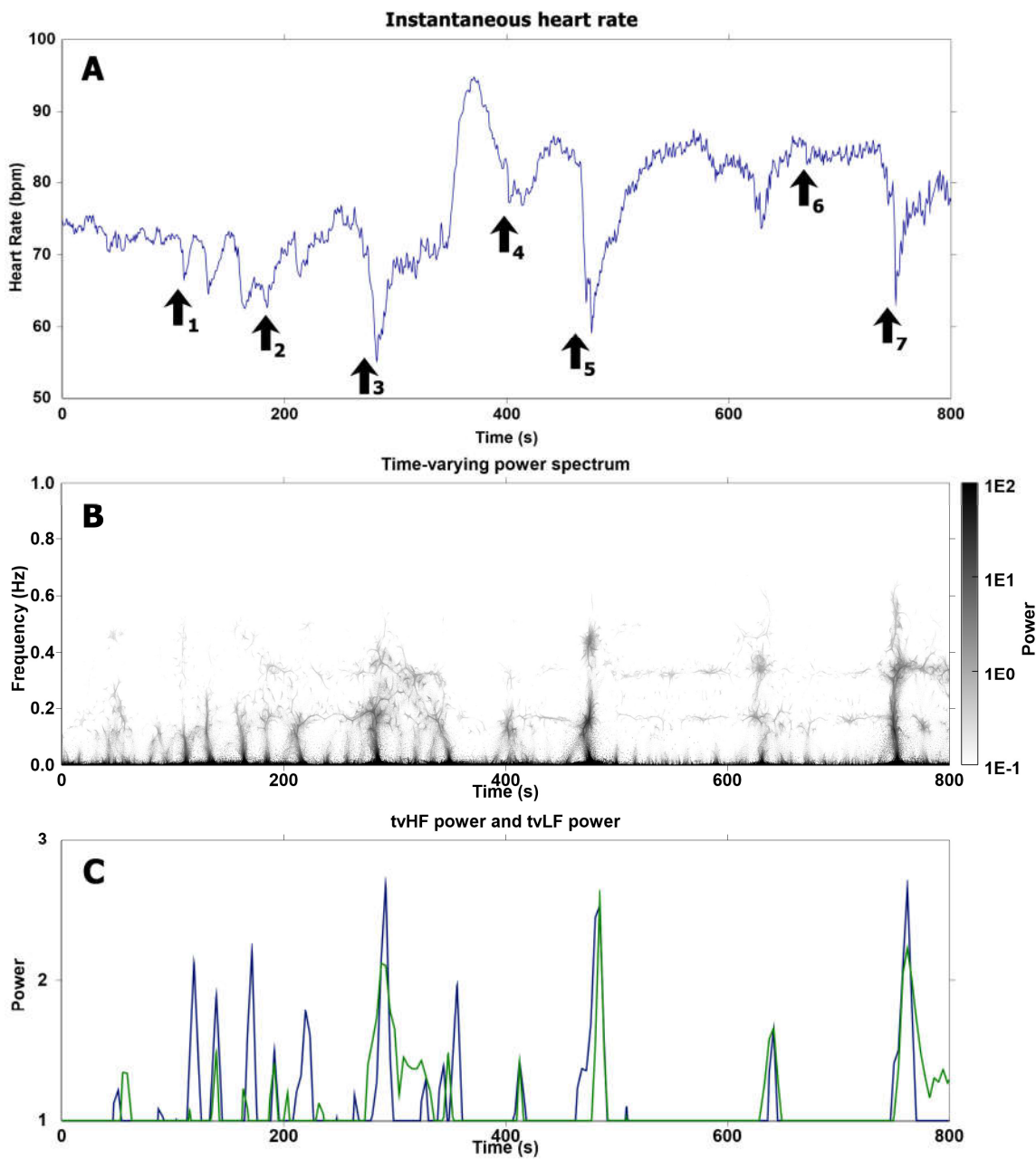

**Fig. 1.** A representative data shows the standard procedure of the laparoscope cholecystectomy and the dynamic IHR changes in panel A: Arrow 1, umbilical skin incision; Arrow 2, Veress needle insertion; Arrow 3, umbilical trocar insertion; Arrow 4, xiphoid skin incision; Arrow 5, xiphoid trocar insertion; Arrow 6, subcostal

skin incision; Arrow 7, subcostal trocar insertion. Panel B shows the corresponding changes in time-varying power spectrum. Panel C shows the corresponding indices, time-varying high frequency (tvHF) power as green colour and time-varying low frequency (tvLF) power as blue colour, which also demonstrate the de-sensitization of skin incision, and the relative higher tvHF in all three trocar insertion in contrast to those in skin incision.

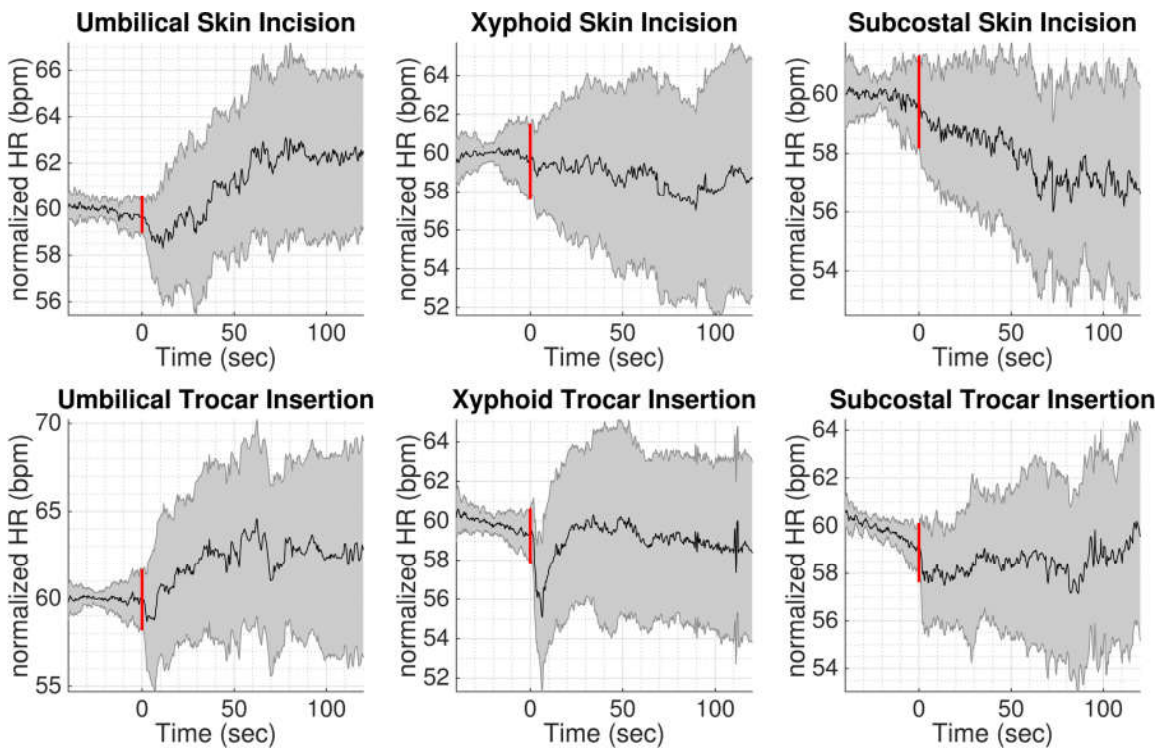

**Fig. 2.** Six sub-figures of group and normalized instantaneous heart rate (IHR) in beat per second (bps) show the de-sensitization of skin incision, and larger transient bradycardia in all three trocar insertion in contrast to that in skin incision. Lines and upper/lower bound represent median and +/– one median absolute deviation respectively.

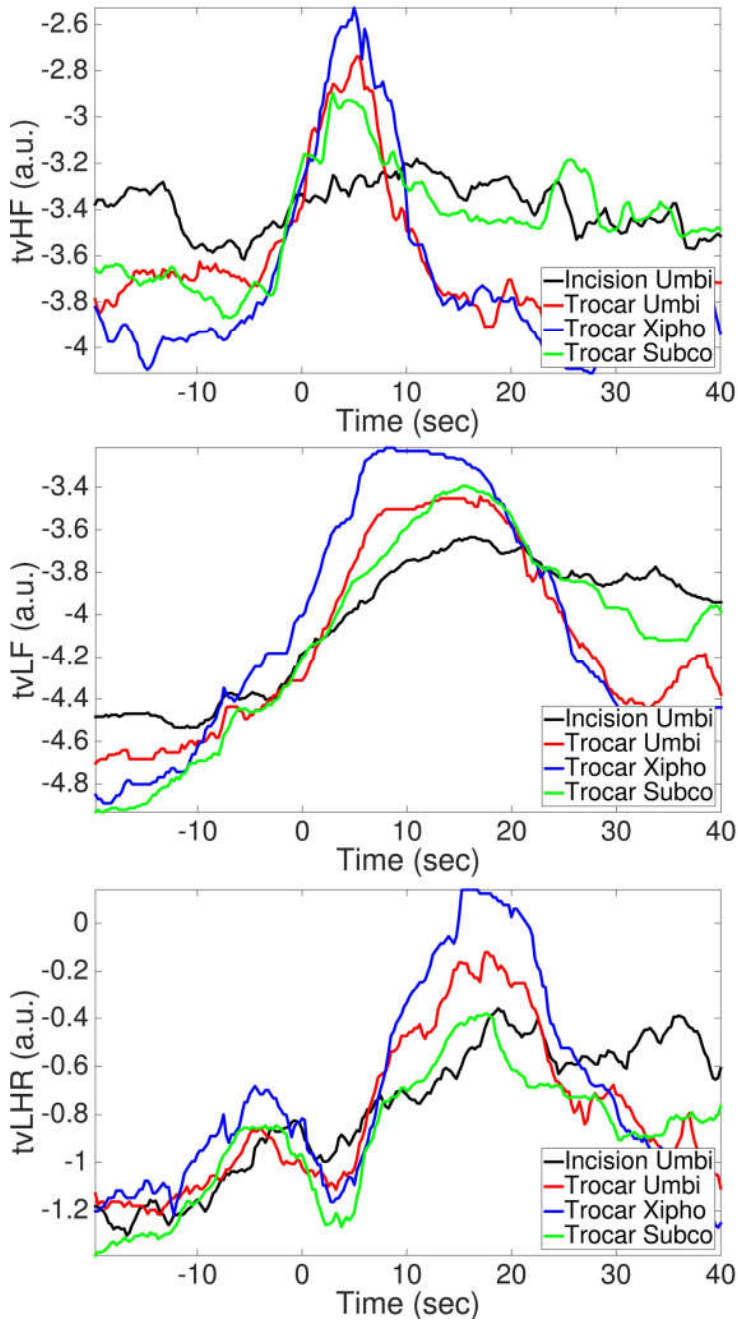

**Fig. 3.** A side-by-side comparison of three indices, time-varying high frequency (tvHF) power, time-varying low frequency (tvLF) power, and tvLF-to-tvHF power ratio (tvLHR). Both time-varying high frequency (tvHF) power and time-varying low frequency (tvLF) power showed that all three trocar insertion elicits significantly stronger autonomic responses than umbilical skin incision. Moreover, trocar insertion elicits more tvHF change than skin incision. The noxious stimuli of trocar insertion are stronger in umbilical and xiphoid area than in subcostal area.

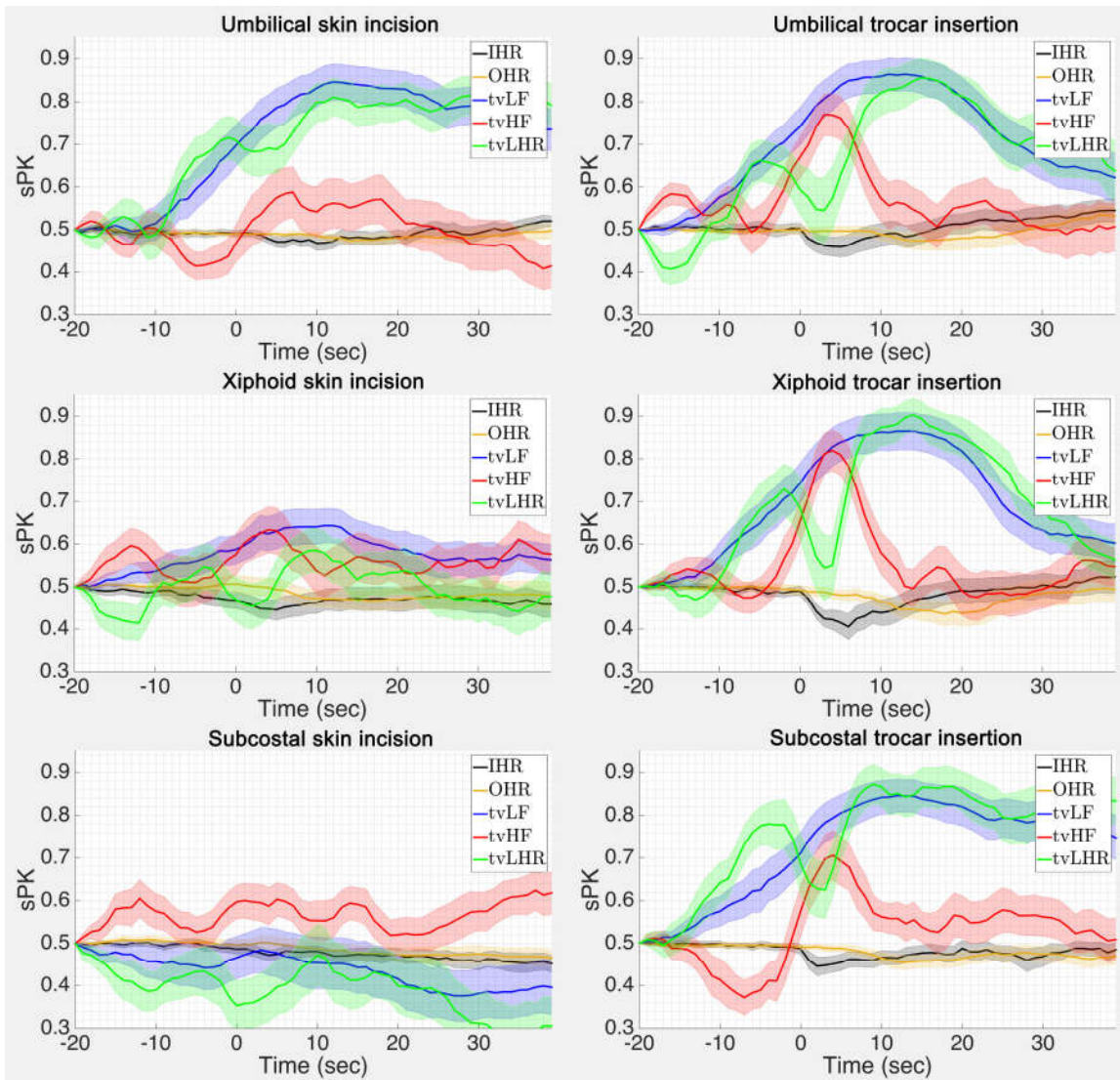

**Fig. 4.** Serial prediction probability (sP$_K$) analyses in six surgical noxious stimuli show the performance as indices and their temporal relationships with respect to each surgical step (zero time). Coloured tracings and faded colour bars represent sP$_K$ value and their standard error respectively. Baseline values are 20 s before surgical steps. The time-varying HF (tvHF) power representing transient bradycardia reaches its peaks faster than time-varying low frequency (tvLF) power and the tvLH-to-tvHF power ratio (tvLHR). The performances of all three indices are better in all three trocar insertion stimuli than in all three skin incision stimuli. The performances of all three indices are better than the ordinary heart rate (OHR ) data from the patient monitor and the instantaneous heart rate (IHR)

in all surgical steps.